\documentstyle{article}

\newcommand{\beqn}{\begin{eqnarray}}
\newcommand{\eeqn}{\end{eqnarray}}
\newcommand{\beqns}{\begin{eqnarray*}}
\newcommand{\eeqns}{\end{eqnarray*}}

\newcommand{\rmd}{\mbox{d}}
\newcommand{\rme}{\mbox{e}}
\newcommand{\rmi}{\mbox{i}}

\newcommand{\dd}[2]{{\rmd{#1}\over\rmd{#2}}}
\newcommand{\pdd}[2]{{\partial{#1}\over\partial{#2}}}
\newcommand{\wt}{\widetilde}

\newcommand{\tr}{\mathop{\mbox{Tr}}}
\newcommand{\var}{\mathop{\mbox{var}}}

\newcommand{\bfn}[1]{\mbox{\protect\bf #1}}
\catcode`\@=11
\newcommand{\ccite}[1]
{\@ifundefined{b@#1}{\bf ?}{\@nameuse{b@#1}}}
\catcode`\@=12

\begin{document}

\title{Quantum Mechanics on a Real Hilbert Space}

\author{
Jan Myrheim\\
Department of Physics, NTNU\\
N--7491 Trondheim, Norway
}

\maketitle

\begin{abstract}
The complex Hilbert space of standard quantum mechanics may be treated
as a real Hilbert space. The pure states of the complex theory become
mixed states in the real formulation. It is then possible to
generalize standard quantum mechanics, keeping the same set of
physical states, but admitting more general observables. The standard
time reversal operator involves complex conjugation, in this sense it
goes beyond the complex theory and may serve as an example to motivate
the generalization. Another example is unconventional canonical
quantization such that the harmonic oscillator of angular frequency
$\omega$ has any given finite or infinite set of discrete energy
eigenvalues, limited below by $\hbar\omega/2$.
\end{abstract}

\section{Introduction}

There are well known mathematical arguments saying that the Hilbert
space of quantum mechanics could be real or quaternionic, as
alternatives to the standard complex theory
\nocite{BvN36}
\nocite{Jauch68}
\nocite{Varadarajan85}
[\ccite{BvN36}--\ccite{Varadarajan85}].
The real case was dealt with mainly by Stueckelberg and collaborators,
who concluded that it is essentially equivalent to the complex case
\nocite{Stueckelberg59}
\nocite{Stueckelberg60}
\nocite{Stueckelberg61a}
\nocite{Stueckelberg61b}
\nocite{Stueckelberg62}
\nocite{Dyson62}
\nocite{Mackey68}
[\ccite{Stueckelberg59}--\ccite{Mackey68}]
(see also \cite{Finkelsteinetal62,Adler95}).
The interest in the quaternionic case is more alive
\nocite{Finkelsteinetal62}
\nocite{Finkelsteinetal63}
\nocite{Adler95}
[\ccite{Finkelsteinetal62}--\ccite{Adler95}].
A more exotic subject is octonionic quantum theory
\cite{Gunaydin73,Gunaydin78}.

The compromise proposed here is a genuine extension of the complex
theory, but is not quite the full quantum theory on a real Hilbert
space. The set of physical states is taken to be exactly the same as
in the complex theory, but the complex Hilbert space is reinterpreted
as a real space, and the set of observables is enlarged from the set
of all complex Hermitean matrices to the set of all real symmetric
matrices.

There exists some physical motivation for such a generalization
in the fact that the time reversal operator $T$ is antilinear in the
complex theory. All transformations involving time reversal are
antilinear, among them the fundamental $CPT$ symmetry of quantum
field theory. It is true that the effect of time reversal can be
described easily enough in standard quantum theory, but strictly
speaking, as soon as time reversal is introduced, the step from the
complex to the real Hilbert space has already been taken
\cite{Dyson62}.

As another example, it is shown in Section~\ref{Harmosc} below how to
represent the canonical Poisson bracket relation $\{x,p\}=1$ in terms
of operators on a finite dimensional real Hilbert space. As is well
known, the canonical commutation relation ${[x,p]}=-\rmi\hbar I$ can
not be represented on a finite dimensional complex Hilbert space,
simply because the commutator on the left hand side must then have
zero trace, while the identity operator $I$ on the right hand side has
nonzero trace. The argument does not apply in the real Hilbert space,
because there the operator $J=\rmi I$ has zero trace.

The main argument of Stueckelberg for the equivalence of real and
complex quantum mechanics is the need for an uncertainty principle.
On the real Hilbert space it is very useful, if not strictly
necessary, to have an operator $J$ commuting with all observables and
having the property that $J^2=-1$, if one wants to derive a general
inequality for the product of the variances of any pair of
observables. However, the argument is not compelling, partly because,
as Stueckelberg points out, there might in principle be one separate
operator $J$ for every pair of observables, and partly because quantum
mechanics makes sense even without a general uncertainty principle. In
the example of Section~\ref{Harmosc} below, the uncertainty principle
for position and momentum holds indeed in all physical states, even
though the position and momentum operators both anticommute with the
operator $J$ defining the complex structure.

Stueckelberg and collaborators also discussed field quantization with
fields that are either linear or antilinear, in the sense of commuting
or anticommuting with $J$. Again their main conclusion is that
quantization with antilinear fields is impossible for bosons, and
possible for fermions but then essentially equivalent to quantization
with linear fields, so that the real case reduces to the complex case.
If the conclusion is valid in the present case, it means that the
unconventional quantization of the harmonic oscillator does not lead
to any interesting new quantum field theory. However, one should
perhaps reexamine the arguments, keeping in mind in particular that
there might be several different square roots of $-1$, as discussed
briefly in Section \ref{Morethanonedf}.

If antilinear field operators are ever going to be useful, it would
most likely be in the quantization of the Dirac field. If the Dirac
matrices are chosen real, then the Dirac equation is seen to be a real
equation, and it is not unnatural to go one step further and formulate
the quantum field theory in terms of a real Hilbert space, with linear
or (possibly?) antilinear field operators, symmetric with respect to
the real scalar product. The standard complex notation, both for the
field equation and for the Hilbert space, hides the fact that the
theory contains several $\rmi=\sqrt{-1}$ that are logically different,
although they are identical by the notation.

One i is the generator of electromagnetic gauge transformations in the
Dirac equation for a charged particle, it commutes with the mass term
in the equation. A second i appears in the massless Weyl equation, it
generates chiral gauge transformations, and does not commute with the
Majorana mass term. For this reason, the Majorana and Dirac mass terms
are claimed to be different, although one might just as naturally have
concluded that there are two different i's and only one kind of mass
term. A third i turns up in the Fourier transformation connecting
the position and momentum representations of the fields, by definition
it commutes with the field operators. The fourth i, acting on the
complex Hilbert space where all the fields act as operators, need not
in principle be identified with any of the three.

\section{Complex and real Hilbert spaces}

For simplicity, we will consider here mostly finite dimensional
Hilbert spaces. The complex Hilbert space $\bfn{C}^D$ of complex
dimension $D$ corresponds to the real Hilbert space $\bfn{R}^{2D}$ of
real dimension $2D$. The imaginary unit i on $\bfn{C}^D$ is then a
linear operator $J$ on $\bfn{R}^{2D}$ with $J^2=-I$. Here $I$ is the
identity operator on $\bfn{R}^{2D}$. We define the correspondence so
that we have for $D=2$, as an example,
\beqn
\psi=\left(\begin{array}{c}
\psi_{1r}+\rmi\psi_{1i}\\ \psi_{2r}+\rmi\psi_{2i}
\end{array}\right)\in\bfn{C}^2
\leftrightarrow
\psi=\left(\begin{array}{c}
\psi_{1r}\\ \psi_{1i}\\ \psi_{2r}\\ \psi_{2i}
\end{array}\right)\in\bfn{R}^4\;.
\eeqn
Then $J$ is an antisymmetric matrix,
\beqn
\label{eq:J4d}
J=\left(\begin{array}{cccc}
0 &-1 & 0 & 0 \\ 
1 & 0 & 0 & 0 \\ 
0 & 0 & 0 &-1 \\ 
0 & 0 & 1 & 0
\end{array}\right).
\eeqn
The complex scalar product on $\bfn{C}^D$,
\beqn
\phi^{\dag}\psi=
\sum_{j=1}^D (\phi_{jr}\psi_{jr}+\phi_{ji}\psi_{ji})+\rmi
\sum_{j=1}^D (\phi_{jr}\psi_{ji}-\phi_{ji}\psi_{jr})\;,
\eeqn
has a real part which is the symmetric scalar product
$\phi^T\psi=\psi^T\phi$ on $\bfn{R}^{2D}$, and an imaginary part which
is the antisymmetric symplectic scalar product
$-\phi^TJ\psi=\psi^TJ\phi$ on $\bfn{R}^{2D}$.

To any complex $D\times D$ matrix $A$ corresponds a real $2D\times 2D$
matrix, which we choose to call by the same name $A$. The
correspondence is such that, for example,
\beqn
%A=
\left(\begin{array}{cc}
A_{11r}+\rmi A_{11i} & A_{12r}+\rmi A_{12i} \\
A_{21r}+\rmi A_{21i} & A_{22r}+\rmi A_{22i}
\end{array}\right)
\leftrightarrow
%A=
\left(\begin{array}{crcr}
A_{11r} &-A_{11i} & A_{12r} &-A_{12i} \\
A_{11i} & A_{11r} & A_{12i} & A_{12r} \\
A_{21r} &-A_{21i} & A_{22r} &-A_{22i} \\
A_{21i} & A_{21r} & A_{22i} & A_{22r}
\end{array}\right)\;.
\eeqn
In tensor product notation we may write
\beqn
A=\left(\begin{array}{cc}
1 & 0\\
0 & 1
\end{array}\right)\otimes
\left(\begin{array}{cc}
A_{11r} & A_{12r} \\
A_{21r} & A_{22r}
\end{array}\right)+
\left(\begin{array}{cc}
0 &-1\\
1 & 0
\end{array}\right)\otimes
\left(\begin{array}{cc}
A_{11i} & A_{12i} \\
A_{21i} & A_{22i}
\end{array}\right)\;.
\eeqn
The Hermitean conjugate $A^{\dag}$ of the complex matrix $A$
corresponds to the transposed $A^T$ of the real matrix $A$. The
distinguishing property of those real matrices that correspond to
complex matrices, is that they commute with $J$. Any real $2D\times
2D$ matrix $A$ can be written in a unique way as $A=A_++A_-$, where
$A_+J=JA_+$ and $A_-J=-JA_-$, in fact the explicit solution is
\beqn
A_{\pm}={1\over 2}\,
(A\mp JAJ)\;.
\eeqn
$A_+$ is complex linear. $A_-$ is complex antilinear, thus it is a
product of complex conjugation and a complex linear operator. In the
$4D^2$ dimensional space of all real matrices, the complex linear and
the complex antilinear matrices form two complementary subspaces of
complex dimension $D^2$ and real dimension $2D^2$.

The complex $D\times D$ matrix $A$ is Hermitean if $A^{\dag}=A$ and
unitary if $A^{\dag}=A^{-1}$. The real $2D\times 2D$ matrix $A$ is
symmetric if $A^T=A$, antisymmetric if $A^T=-A$, orthogonal if
$A^T=A^{-1}$, and symplectic if $A^TJ=JA^{-1}$. An orthogonal matrix
is symplectic if and only if it commutes with $J$. Thus, the complex
Hermitean matrices correspond to those real matrices that are
symmetric and commute with $J$, whereas the complex unitary matrices
correspond to precisely those real matrices that are orthogonal and
symplectic.

In other words, an orthogonal matrix is a real linear operator on
$\bfn{R}^{2D}$ which is invertible and preserves the ordinary real
scalar product $\phi^T\psi$. A symplectic matrix is invertible and
preserves the symplectic scalar product $-\phi^TJ\psi$. And a unitary
matrix is a complex linear operator on $\bfn{C}^D$ which is invertible
and preserves the complex scalar product $\phi^{\dag}\psi$. (In the
finite dimensional case, but not in the infinite dimensional case, the
invertibility is a consequence of the preservation of scalar
products.)

An infinitesimal linear transformation $U=I+\epsilon A$ on
$\bfn{R}^{2D}$, with $\epsilon$ infinitesimal, is orthogonal if and
only if the generator $A$ is antisymmetric, $A^T=-A$. It is
symplectic if and only if $A^TJ=-JA$, which means that the matrix
$B=JA$ is symmetric. Equivalently, $A=-JB$ with $B$ symmetric. $U$
is both orthogonal and symplectic if and only if $A=-JB=-BJ$ with $B$
symmetric.

The dimension of the orthogonal group $\mbox{O}(2D)$ is $2D^2-D$, the
dimension of the symplectic group $\mbox{Sp}(2D)$ is $2D^2+D$, and the
dimension of the unitary group $\mbox{U}(D)$ is $D^2$.

\pagebreak[4]
\section{Quantum Mechanics}

\subsection{Observables and probabilities}

In standard quantum mechanics a pure state of a given physical system
is represented by a unit vector $\psi\in\bfn{C}^D$, or equivalently by
the Hermitean density matrix $\rho=\psi\psi^{\dag}$, which is a
projection operator, since $\rho^2=\rho$. An observable of the system
is represented by a Hermitean $D\times D$ matrix $A$, and the theory
predicts the expectation value in the pure state $\psi$ as
\beqn
\langle A\rangle
=\psi^{\dag}A\psi=\tr(\rho A)\;.
\eeqn
More generally, any pure or mixed state is represented by a Hermitean
density matrix $\rho$ which is positive definite, i.e.\ has nonnegative
eigenvalues, and has unit trace, $\tr\rho=1$. The expectation value of
the observable $A$ in this state is $\langle A\rangle=\tr(\rho A)$.
The theory also predicts the variance of $A$, $\var(A)=(\Delta A)^2$,
as
\beqn
\var(A)
=\langle(A-\langle A\rangle)^2\rangle
=\langle A^2\rangle-\langle A\rangle^2\;.
\eeqn
$A$ has a sharp value in the state $\rho$ if and only if $\var(A)=0$.

This probability interpretation makes sense because of the spectral
theorem for Hermitean matrices, which guarantees the existence of the
spectral representation
\beqn
A=\sum_{n=1}^N a_nP_n\;.
\eeqn
Here $a_1,\ldots,a_N$ are distinct real eigenvalues, $N\leq D$, and
$P_1,\ldots,P_N$ are Hermitean projection operators, with the
properties that $P_mP_n=0$ for $m\neq n$, and
\beqn
I=\sum_{n=1}^N P_n\;.
\eeqn
This implies that
\beqn
\langle A\rangle=\sum_{n=1}^N p_na_n\;,\qquad
\var(A)=\sum_{n=1}^N p_n(a_n-\langle A\rangle)^2\;,
\eeqn
where $p_n=\langle P_n\rangle=\tr(\rho P_n)$. According to the
probability interpretation, the possible results of a measurement of
$A$ are the eigenvalues $a_1,\ldots,a_N$, and $p_n$ is the probability
of the result $a_n$ in the state $\rho$.

The fact that the spectral theorem for complex Hermitean matrices is
valid for all real symmetric matrices, with no more than the obvious
changes in wording, allows us to generalize standard quantum mechanics
by admitting as observables all the real symmetric matrices.

\subsection{States}

In this generalization we have two options for choosing the set of
states. The straightforward choice is to admit all real unit vectors
as possible pure states of the system, and all real symmetric and
positive definite matrices of unit trace as possible mixed states. This
enlarges the set of possible states as compared to standard quantum
mechanics, since not all such real density matrices are complex
linear. In particular, it doubles the total number of states in the
system, and it doubles the degeneracy of the spectrum of all standard
observables, i.e.\ those that are complex linear. It seems that the
degeneracy doubling is unphysical, at least if we want to describe
systems that are well described by standard quantum theory.

The second option, more interesting from the physical point of view,
is to admit exactly the same states as in the complex theory. It means
that we enlarge the class of observables, including all real symmetric
matrices, but we admit only those density matrices that commute with
$J$. For example, with $J$ as in equation (\ref{eq:J4d}) the most
general physical density matrix has the form
\beqn
\label{eq:rho4d}
\rho=\left(\begin{array}{cccc}
\alpha & 0 & \gamma & \delta \\
0 & \alpha &-\delta & \gamma \\ 
\gamma &-\delta & \beta & 0 \\
\delta & \gamma & 0 & \beta
\end{array}\right),
\eeqn
with $2(\alpha+\beta)=1$, $\alpha\geq 0$, $\beta\geq 0$, and
$\alpha\beta-\gamma^2-\delta^2\geq 0$.

This has the advantage that the physical degeneracies are unchanged,
but it also has the somewhat strange consequence that there are no
pure states in the theory. In fact, the pure states in the complex
theory correspond to mixed states in the real theory. One way to see
this is to observe that the meaning of the trace is different in the
real theory as compared to the complex theory, because the number of
basis vectors is doubled. Thus, if $\tr\rho=1$ when the density matrix
$\rho$ is regarded as a complex $D\times D$ matrix, we have
$\tr\rho=2$ when the same $\rho$ is regarded as a real $2D\times 2D$
matrix. Therefore the proper correspondence is that the complex
density matrix $\rho$ must correspond to the real density matrix
$\rho/2$, which can never represent a pure state since it has no
eigenvalues larger than $1/2$.

In this generalization of quantum mechanics as a theory defined on
$\bfn{R}^{2D}$, every observable will have a complete set of $2D$ real
eigenvalues and eigenvectors. However, one eigenvector alone does not
represent a physical state. If we say that two physical states $\rho$
and $\rho'$ are orthogonal when $\rho\rho'=\rho'\rho=0$, the maximum
number of orthogonal physical states is $D$. Just by counting we see
that if an observable $A$ has more than $D$ different eigenvalues, not
every one of these can possess its own ``physical eigenstate'', in
which a measurement of $A$ gives this particular value with
probability one.

More explicitly stated, if an observable $A$ does not commute with
$J$, then it will have at least one eigenvalue which is not sharply
realized in any physical state, and conversely, there will exist no
complete set of physical states such that $\var(A)=0$ in every state
belonging to the complete set.

\subsection{Poisson brackets}

To the classical Poisson bracket $\{A,B\}$ of two classical
observables $A$ and $B$ corresponds the ``quantum Poisson bracket''
\beqn
\{A,B\}
=-{\rmi\over\hbar}\,{[A,B]}
=-{\rmi\over\hbar}\,(AB-BA)\;.
\eeqn
The proper way to write the same quantity in the real formulation is
\beqn
\{A,B\}
=A\Omega B-B\Omega A\;,
\eeqn
with $\Omega=-J/\hbar$. The antisymmetry, $\{A,B\}=-\{B,A\}$, and the
Jacobi identity,
\beqn
 \{A,\{B,C\}\}
+\{B,\{C,A\}\}
+\{C,\{A,B\}\}=0\;,
\eeqn
are easily verified. The most important property of the matrix
$\Omega$ is that it is antisymmetric, $\Omega^T=-\Omega$, because that
ensures that $\{A,B\}$ is symmetric whenever $A$ and $B$ are both
symmetric. Another way to write the relation $\{A,B\}=C$ for symmetric
matrices $A$, $B$ and $C$ is as
\beqn
{[-JA,-JB]}=-\hbar JC\;.
\eeqn
This is then a commutation relation in the Lie algebra of the
symplectic group.

The real density matrix $\rho$ must in general be explicitly time
dependent and satisfy the Liouville equation
\beqn
\dd{\rho}{t}=
\pdd{\rho}{t}+\{\rho,H\}
=0\;.
\eeqn
Here $\rmd\rho/\rmd t$ is the absolute time derivative,
$\partial\rho/\partial t$ is the explicit time derivative, and $H$ is
the Hamiltonian. Thus the explicit time dependence of $\rho$ is given
by the equation of motion
\beqn
\pdd{\rho}{t}=\{H,\rho\}
=H\Omega\rho-\rho\Omega H\;.
\eeqn
The equation of motion must preserve $\tr\rho$. A sufficient
condition is that either $H$ or $\rho$ commute with $\Omega$, because
then we have either $\rho\Omega H=\rho H\Omega$ or $\rho\Omega
H=\Omega\rho H$, and in both cases
\beqn
\pdd{(\tr\rho)}{t}
=\tr(H\Omega\rho-\rho\Omega H)=0\;.
\eeqn
Thus, if we accept all symmetric and positive definite matrices of unit
trace as density matrices, we should impose the condition on the
Hamiltonian $H$ that it commute with $J=-\hbar\Omega$.

We should impose the same condition on $H$ even in the case where we
accept only density matrices that commute with $J$. The point is that
the equation of motion must preserve the condition of commutation with
$J$, that is, the Poisson bracket $\{H,\rho\}= H\Omega\rho-\rho\Omega
H$ must commute with $J$. A sufficient condition, when $\rho$ commutes
with $J$, is that $H$ also commutes with $J$.

When $H$ commutes with $J$, and is not explicitly time dependent, the
equation of motion can be integrated explicitly to give
\beqn
\rho(t)=U(t)\,\rho(0)\,U(-t)\;,
\eeqn
where $U(t)=\rme^{-{t\over\hbar}\,JH}$ is the unitary time development
operator.

In conclusion, not every real symmetric matrix is an acceptable
Hamiltonian in the generalized quantum mechanics as formulated
here. It is necessary, or at least natural, to require the Hamiltonian
to be a complex linear matrix. If we also require the density matrices
to be complex linear matrices, it would seem that we are back to the
point of departure, which was the standard complex quantum
mechanics. However, we have enlarged the class of observables, even
though we do not accept the new observables to be Hamiltonians
governing the time development.

\section{The harmonic oscillator}
\label{Harmosc}

Time reversal was mentioned in the introduction as a motivation for
the proposed generalization of the complex formalism. A more
unconventional example of the generalizations that become possible, is
the representation of the canonical Poisson bracket $\{x,p\}=I$ in any
finite and even dimension. For example, in the four dimensional case
considered above, any two positive lengths $\xi_1,\xi_2$ define a
representation of the form
\beqn
\label{xpfourdim}
x=\left(\begin{array}{cccc}
\xi_1 & 0 & 0 & 0 \\ 
0 &-\xi_1 & 0 & 0 \\ 
0 & 0 &\xi_2 & 0 \\ 
0 & 0 & 0 &-\xi_2
\end{array}\right),\qquad
p={\hbar\over 2}\left(\begin{array}{cccc}
0 & 1/\xi_1 & 0 & 0 \\ 
1/\xi_1 & 0 & 0 & 0 \\ 
0 & 0 & 0 &1/\xi_2 \\ 
0 & 0 &1/\xi_2 & 0
\end{array}\right).
\eeqn
The generalization to any even dimension $2D$, or to infinite
dimension, is obvious. Then both $x$ and $p$ anticommute with the
imaginary unit $J$, as opposed to standard quantum mechanics where
they commute with $J$. With the most general physical density matrix
$\rho$, equation (\ref{eq:rho4d}), this gives $\langle x\rangle=0$,
$(\Delta x)^2=\langle x^2\rangle
=2(\alpha\xi_1^{\;2}+\beta\xi_2^{\;2})$, and similarly for $p$. Thus
the Heisenberg uncertainty relation holds in every physical state,
\beqn
\Delta x\,\Delta p
&\!\!\!=&\!\!\!
\hbar\,\sqrt{(\alpha\xi_1^{\;2}+\beta\xi_2^{\;2})
\left({\alpha\over\xi_1^{\;2}}
+{\beta\over\xi_2^{\;2}}\right)}
\nonumber\\
&\!\!\!=&\!\!\!
\hbar\,\sqrt{(\alpha+\beta)^2+\alpha\beta\left(
{\xi_1\over\xi_2}-{\xi_2\over\xi_1}\right)^2}
\geq{\hbar\over 2}\;.
\eeqn

With these definitions the Hamiltonian of the harmonic oscillator of
angular frequency $\omega$,
\beqn
\label{hoHam}
H={p^2\over 2m}+{1\over 2}\,m\omega^2x^2\;,
\eeqn
is diagonal and has the energy eigenvalues
\beqn
E_i={\hbar^2\over 8m\xi_i^{\;2}}
+{1\over 2}\,m\omega^2\xi_i^{\;2}\;.
\eeqn
For any given value $E_i\geq\hbar\omega/2$ this equation has the
positive solutions
\beqn
\xi_i=\sqrt{2E_i\pm
\sqrt{4E_i^{\;2}-\hbar^2\omega^2}\over 2m\omega^2}\;.
\eeqn
By the obvious generalization, we may assign to the harmonic
oscillator any finite number, or an infinite number, of arbitrary
energy levels above the lower bound $\hbar\omega/2$.

Note that the Hamiltonian $H$ does commute with $J$, even though $x$
and $p$ here do not, implying that the time development operator
$\rme^{-{t\over\hbar}\,JH}$ is unitary. On the other hand, the
operator $\rme^{-{d\over\hbar}\,Jp}$, representing translation in
space by a distance $d$, is not unitary when $d\neq 0$, but only
symplectic. The problem is that it is not orthogonal, because the
matrix $Jp$ in the exponent is symmetric rather than antisymmetric.
Thus it does not conserve probabilities, and is not a symmetry
transformation, as is evident from the fact that the position
operator $x$ has a discrete spectrum.

The case $\xi_1=\xi_2=\xi$ in equation (\ref{xpfourdim}) is
interesting because it is a realization of the canonical relation
$\{x,p\}=I$, but is also, in a certain sense, a
fermionic quantization of the harmonic oscillator, with
\beqn
x^2=\xi^2\;,\qquad
p^2={\hbar^2\over 4\xi^2}\;,\qquad
xp+px=0\;.
\eeqn
In fact, the last relations are just one particular form of the
canonical anticommutation relations
\beqn
aa^T+a^Ta=I\;,\qquad
a^2=(a^T)^2=0\;.
\eeqn
To see this in more detail, let us introduce another antisymmetric
matrix $K$ so that it is an imaginary unit,
$K^2=-I$, and commutes with both $x$ and
$p$. Then we take
\beqn
a  ={1\over 2\xi}\,x+{\xi\over\hbar}\,Kp\;,\qquad
a^T={1\over 2\xi}\,x-{\xi\over\hbar}\,Kp\;.
\eeqn
The Hamiltonian $H$ as defined in equation (\ref{hoHam}) is
just a constant, and a more interesting quantity is the usual
Hamiltonian of the fermionic oscillator, which is
\beqn
H'=\hbar\omega\left(a^Ta-{1\over 2}\right)
=-{\omega\over 2}\,K(xp-px)
={\hbar\omega\over 2}\,JK\;.
\eeqn
The time development operator with this alternative Hamiltonian is
\beqn
U'(t)=\rme^{-{t\over\hbar}\,JH'}=\rme^{{\omega t\over 2}\,K}\;.
\eeqn

In some respects this theory, where $x$ and $p$ anticommute with the
imaginary unit $J$, is equivalent to another theory where the
imaginary unit is $K$, which commutes with $x$ and $p$. An explicit
representation for $K$ might be
\beqn
K=\left(\begin{array}{cccc}
0 & 0 &-1 & 0 \\ 
0 & 0 & 0 &-1 \\ 
1 & 0 & 0 & 0 \\ 
0 & 1 & 0 & 0
\end{array}\right)=SJS^{-1}\;,
\eeqn
with, for example,
\beqn
S=S^{-1}=\left(\begin{array}{cccc}
1 & 0 & 0 & 0 \\ 
0 & 0 & 1 & 0 \\ 
0 & 1 & 0 & 0 \\ 
0 & 0 & 0 & 1
\end{array}\right)\;.
\eeqn
If we define
\beqn
\wt{\rho}=S\rho S^{-1}
=\left(\begin{array}{cccc}
\alpha & \gamma & 0 & \delta \\
\gamma & \beta &-\delta & 0 \\ 
0 &-\delta & \alpha & \gamma \\
\delta & 0 & \gamma & \beta
\end{array}\right),
\eeqn
then this is a density matrix which commutes with $K$ instead of with
$J$. If now
$\rho(t)=U'(t)\,\rho(0)\,U'(-t)$,
then we have
$\wt{\rho}(t)=\wt{U}(t)\,\wt{\rho}(0)\,\wt{U}(-t)$,
where
\beqn
\wt{U}(t)=S\,U'(t)\,S^{-1}=\rme^{-{t\over\hbar}\,KH'}
=\rme^{{\omega t\over 2}\,J}\;,
\eeqn
since $SH'=H'S$. The energy spectrum is the same in the two theories,
for example we have
\beqn
\tr(\rho H')=\tr(\wt{\rho}H')=2\delta\hbar\omega\;.
\eeqn
However, the expectation values for $x$ and $p$ are not the same.
For example,
\beqn
\tr(\rho x)=0\;,\qquad
\tr(\wt{\rho} x)=2(\alpha-\beta)\xi\;.
\eeqn

\section{More than one degree of freedom}
\label{Morethanonedf}

With two degrees of freedom, referred to here by indices $a$ and $b$,
and belonging for example to two fields commuting with each other, the
Hilbert space is a tensor product,
\beqn
{\cal H}={\cal H}_a\otimes{\cal H}_b\;.
\eeqn
However, the complex and real tensor products are mathematically
different constructions. The complex tensor product of spaces of
complex dimensions $D_a$ and $D_b$ has complex dimension $D_aD_b$ and
real dimension $2D_aD_b$, whereas the real tensor product of spaces of
real dimensions $2D_a$ and $2D_b$ has dimension $4D_aD_b$.

The relation between the two types of tensor product can be understood
as follows. In the complex case the following relations hold for
tensor products of vectors,
\beqn
\phi\otimes\psi
=-\rmi((\rmi\phi)\otimes\psi)
=-\rmi(\phi\otimes(\rmi\psi))
=-(\rmi\phi)\otimes(\rmi\psi)\;.
\eeqn
In the real case the four tensor products $\phi\otimes\psi$,
$(J\phi)\otimes\psi$, $\phi\otimes(J\psi)$ and $(J\phi)\otimes(J\psi)$
are linearly independent vectors, thus there exist two imaginary units
$J_a=J\otimes I$ and $J_b=I\otimes J$ defined as operators on the real
tensor product space ${\cal H}$. It follows that ${\cal H}$ is a direct
sum, ${\cal H}={\cal H}_+\oplus{\cal H}_-$, where
${\cal H}_{\pm}=P_{\pm}{\cal H}$, and the two operators
\beqn
P_{\pm}={1\over 2}\,(I\mp J_aJ_b)
\eeqn
are complementary orthogonal projection operators. Each of the two
subspaces ${\cal H}_{\pm}$ has dimension $2D_aD_b$. Both $J_a$ and
$J_b$ commute with $P_{\pm}$, and hence act within the subspaces
${\cal H}_{\pm}$ separately. On ${\cal H}_+$ the relation $J_a=J_b$
holds, whereas $J_a=-J_b$ holds on ${\cal H}_-$.

The complex tensor product space can be identified in a natural way
with ${\cal H}_+$. Hence a physical density matrix $\rho$ on the
product space ${\cal H}$ must be a real $(4D_aD_b)\times(4D_aD_b)$
matrix that commutes with both $J_a$ and $J_b$, and in addition it
must have the property that $\rho=P_+\rho=\rho P_+=P_+\rho P_+$.

If an operator $x$ acts on ${\cal H}_a$ and anticommutes with $J$,
then the corresponding operator $x_a=x\otimes I$ on
${\cal H}={\cal H}_a\otimes{\cal H}_b$
anticommutes with $J_a$, but commutes with $J_b$. This means that
$x_a$ maps ${\cal H}_+$ into ${\cal H}_-$, and vice versa. Such an
operator could not be constructed at all within the complex quantum
theory, because the subspace ${\cal H}_-$ simply would not exist. It
is essential for the construction that there exist two mutually
commuting imaginary units $J_a$ and $J_b$ on the real tensor product
Hilbert space. The possibility of having operators that are antilinear
with respect to $J_a$ and at the same time linear with respect to
$J_b$, or vice versa, is at least a partial answer to one of the
problems with antilinear field operators recognized by Stueckelberg
and collaborators.

The operator $x_a$ above is ``unphysical'' in the sense that it maps
the physical space ${\cal H}_+$ out of itself. But similar unphysical
operators are well known in physics. For example, a general isospin
rotation does not respect the superselection rule for electric charge,
it transforms a physical state into an unphysical superposition of
states with different values of the charge. Another example is the
relative position of two identical particles, which is an operator
changing the symmetry properties of the two-particle wave functions.

The generalization to more than two factors in the tensor product is
easy, just introduce one new pair of projection operators for each new
factor. If for example ${\cal H}={\cal H}_a\otimes{\cal
H}_b\otimes{\cal H}_c$, then define $P_{\pm}$ as above, and in
addition
\beqn
Q_{\pm}={1\over 2}\,(I\mp J_aJ_c)\;.
\eeqn
Since $P_{\epsilon}$ and $Q_{\eta}$ commute, for $\epsilon=\pm$ and
$\eta=\pm$, the products $P_{\epsilon}Q_{\eta}$ are also projection
operators. Decompose ${\cal H}$ as a direct sum of subspaces
${\cal H}_{\epsilon\eta}=P_{\epsilon}Q_{\eta}{\cal H}$, then in each
such subspace we have $J_a=\epsilon J_b=\eta J_c$. The physical
subspace is ${\cal H}_{++}$, it corresponds to the complex tensor
product, since the relations $J_a=J_b=J_c$ hold there.

\end{document}